# SPATIALLY RESOLVED ANALYSES OF MICROWAVE AND INTERMODULATION CURRENT FLOW ACROSS HTS RESONATOR USING LOW TEMPERATURE LASER SCANNING MICROSCOPY

Alexander P. Zhuravel
B. Verkin Institute for Low temperature Physics & Engineering, NAS of Ukraine
Address: 47 Lenin Avenue, Kharkov, 61103, Ukraine
Tel.: +380-572-308507, Fax: +380-572-322370, E-mail: zhuravel@ilt.kharkov.ua
Stephen Remillard
Agile Devices
Address: 906 University Place, Evanston, IL 60201-3121 USA
Tel.: 847-570-4392, Fax: 847-866-1808, E-mail: remillard@agile-devices.com
Steven M. Anlage
Physics Department, Center for Superconductivity Research, University of Maryland
Address: College Park, MD 20742-4111 USA
Tel.: (301) 405-7321, Fax: (301) 405-3779, E-mail: anlage@squid.umd.edu
Alexey V. Ustinov
Physics Institute III, University of Erlangen-Nuremberg
Address: Erwin-Rommel Str. 1, D-91058, Erlangen, Germany
Tel.: +49 9131 85 27268, Fax: +49 9131 15249, E-mail: ustinov@physik.uni-erlangen.de

Microwave resonators made from high-temperature superconducting (HTS) films are becoming increasingly commercially competitive in the market of modern communication technologies. However, unlike normal metals, the HTS device performance suffers from intrinsically nonlinear electrodynamics, resulting in harmonic generation and intermodulation (IM) distortion at all levels of circulating power. The dominant sources of IM response are believed to be inductive and localized in only small regions of the HTS film having the highest current densities $J_{RF}(x,y)$. In our previous papers, examining microwave resonators of simple microstrip geometry, we have demonstrated that local nonlinearity imaging is possible with the technique of low temperature laser scanning microscopy (LTLSM) adapted to microwave experiments. Unique contrast due to IM current density $J_{IM}(x,y)$ was generated by the IM imaging method. It was applied to display the regions showing large nonlinear LTLSM photoresponse (PR) in HTS devices with a few-micrometer spatial resolution [1-3]. Direct local correlation between $J_{IM}(x,y)$ and high $J_{RF}(x,y)$ at the edges of the strips, typical defects like twin-domain blocks, in-plane rotated grains and micro-cracks was shown. Here we want to indicate that the geometrical features in HTS circuit layout for microwave devices of more complex geometry also radically affect the linear device performance.

Fig. 1(a) shows the geometry of the RF resonator under test. The device represents a transmission line patterned from thin $Tl_2Ba_2CaCu_2O_y$ film on $LaAlO_3$ substrate into a meandering strip by ion-milling lithography. It is capacitive coupled to an input RF circuit delivering power $P_{IN}$ in the range from –40 to +10 dBm. The frequency of fundamental resonance is about 1.85 GHz with a loaded $Q_L\sim 2000$ at T=80 K. For LTLSM characterization, the resonator was placed inside the vacuum cavity of a variable-temperature optical cryostat which stabilizes the temperature of the sample in the range *6-300* K with an accuracy of *5* mK. Under this operating condition, the sample was *x-y* scanned by a 4 μm diameter thermal probe produced by a 1.1 μm diameter focused laser beam that is amplitude modulated with a frequency of about 100 kHz on the sample surface. The oscillating probe excites PR=$\delta V(x,y)$ that was detected by a spectrum analyzer, amplified by lock-in techniques, and then used for plotting the spatio-amplitude map of a quantity proportional to $J_{RF}(x,y)$ squared as a function of position *(x,y)* of the laser focus on the sample. Fig. 1(b) represents a gray-scale LTLSM image showing the spatial variation of $J_{RF}(x,y)$ squared in the resonator at the third harmonic resonance (5.9 GHz), T=83 K, and $P_{IN}$ =0 dBm. Brighter areas in the image correspond to higher *rf* current densities. As can be seen from the image, the distribution of $J_{RF}(x,y)$ forms a typical standing wave pattern having peak current densities at the edges of HTS strip line. For clarity, we reconstructed the LTLSM image into an amplitude profile of $J_{RF}(x,y)$ along the longitudinal path from A to H constituting the length of the strip. To make this profile [see Fig. 1(c)], data for $J_{RF}(x,y)$ were averaged in each cross-section of the strip. The best fit to the plot [solid line in Fig. 1(c)] is $\cos^2(3kz+\phi)$, as expected for $J_{RF}(x,y)$ squared in a straight one-dimensional resonator. It is not a surprise that the standing wave pattern is distorted in the A-C and F-H regions. It is possible to explain this behavior as a result of interference of the large screening currents at the edges of two nearby conductors carrying equal and opposite currents in the center of the resonator. The dominant feature of $J_{RF}(x,y)$ is the abnormally high RF PR seen in a few isolated areas (in the vicinity of B, C, D, E, F and G) on the inner corners of the HTS structure. These areas of huge RF current densities are expected to be the principal candidates for generation of local sources of nonlinearity in the device and were studied in detail by using the LTLSM in IM imaging mode.

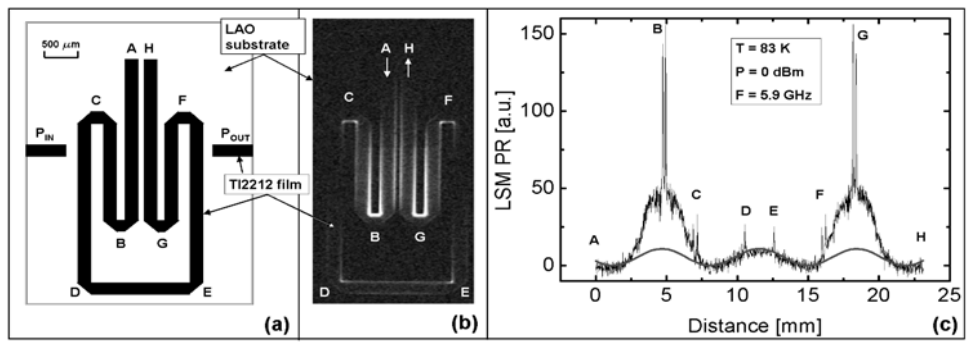

Fig. 1. (a) Top view of the resonator geometry, (b) LTLSM image and (c) profile of standing wave pattern

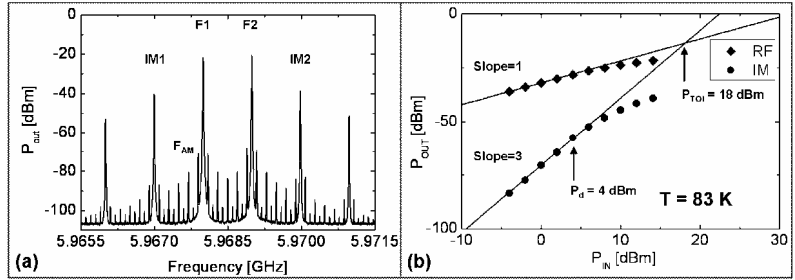

Fig. 2. (a) Spectrum of output signals at $P_{IN}$ = +14 dBm and (b) power dependence of global RF/IM response of the resonator used for IM PR imaging.

Fig. 2 shows *global* characteristics of the resonator used to make IM images. A set of $J_{IM}(x,y)$ LTLSM images was obtained at the IM2 frequency [Fig.2(a)] resulting from nonlinear mixing of the two primary tones (F1 = 5.968 GHz and F2 = 5.969 GHz) of equal power $P_{IN}$ that were centered with 1 MHz spacing around the peak frequency F = 5.9685 GHz of the device third harmonic. The spectrum was obtained at T=83 K and $P_{IN}$ = +14 dBm. The solid points in Fig.2(b) indicate the entire set of $P_{IN}$ values used for LTLSM measurements. To estimate the amplitudes of the linear ($\delta V_{RF}(x,y)$) and nonlinear ($\delta V_{IM}(x,y)$) photoresponse, the laser beam was positioned at the corner of the resonator structure having a maximum value of $J_{RF}(x,y)$. As can be seen from the $F_{AM}$ signal in Fig.2 (a), the 100 kHz modulation of the probe produces frequency modulated microwave response on the order of $10^{-5}$-$10^{-6}$ of the input CW signals.

Fig. 3(a) shows an optical reflectivity LTLSM map (upper image) and the corresponding 3D plot (at bottom) of $J_{IM}(x,y)$ for a 0.4x0.4 mm$^2$ area of the resonator that manifested the highest amplitude of $J_{RF}(x,y)$. As evident, both the linear $J_{RF}(x,y)$ and nonlinear $J_{IM}(x,y)$ components of the LTLSM PR are peaked at the same positions. The power-dependent spatial redistribution of these responses was analyzed in detail in a 25x25 $\mu m^2$ region that is shown in Fig. 3(a) by a white open box containing an inside corner. No visible change in the $J_{RF}(x,y)$ distribution was detected for input powers between -40 dBm and +10 dBm at all temperatures from 78 K to $T_c$. A typical LSM image is presented in Fig. 3(b). In contrast, the IM PR undergoes a radical redistribution starting from -4 dBm here, especially for $P_{IN}$ close to +4 dBm [Fig. 3(c-e)] when the barely detectable 8 dBm IMD is visible in the global microwave response [see Fig. 2(b)].

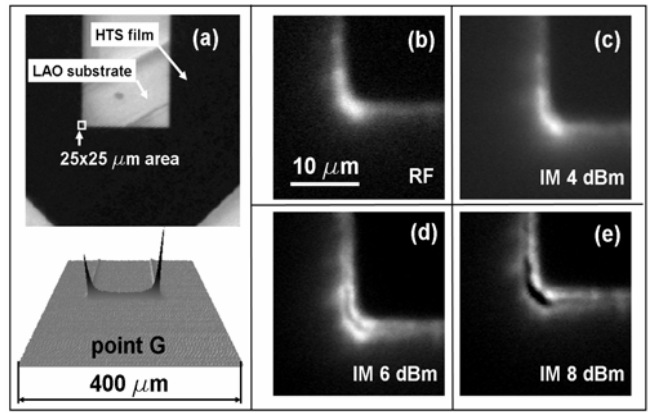

Fig. 3. Spatial correlation of IM and RF photoresponse

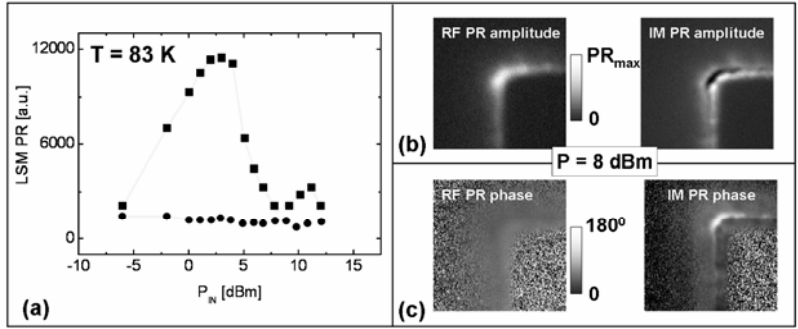

Fig. 4. (a) Power dependence of LSM PR and (b) images of a phase contrast between $J_{RF}(x,y)$ and $J_{IM}(x,y)$

We have positioned a defocused 100 $\mu m$ diameter laser beam at one of the inner corners of the resonator structure to measure the averaged PR data without scanning. Fig. 4(a) shows the power dependence of IM PR (squares) and RF PR (circles) at the corner at T=83 K. As evident, the linear rf PR is practically independent of input rf power. This is due to the use of logarithmic amplification to detect the LSM PR, which presents the PR in the form of $\delta P/P_{IN}$. However, a tremendous change in IM PR is clearly visible in the plot. By using the LTLSM we show that continuous growth of IM PR at small $P_{IN}$ is associated with a widening of the nonlinear current spatial scale, while the decrease of the IM PR is due to a change in amplitude of $J_{IM}$. We note that this drop is associated with the deviation of the *global* third harmonic dependence from power-3 behavior (Fig. 2(b)). Additionally, a novel phase contrast in the LSM PR was found between $J_{RF}(x,y)$ and $J_{IM}(x,y)$. Fig. 4(b,c) compares both the amplitude and phase distribution of these signals obtained at the same T=83 K and P=+8 dBm. It is evident that the phase of $J_{RF}(x,y)$ is practically independent of position while the phase for $J_{IM}(x,y)$ PR changes up to 180 degrees at the corner of the resonator structure. Such behavior is possible when either Abrikosov ($\lambda <$ thickness d) or Pearl ($\lambda > d$) vortices may be generated by high current densities at the corner. These vibrating vortices have a tendency to move into the film under the influence of the Lorenz force with the frequency of the applied rf drive and to create electrical field of opposite sign to the applied one. However, effects of heating in the HTS film by hot-spots of high rf current densities should not to be excluded from consideration as well. The lack of a theory for nonlinear LSM PR does not allow us to make a definitive conclusion. Development of such a theory will be our future effort.

In summary, we have developed the LTLSM technique to search for intrinsic sources of microwave nonlinearities in large area complicated microwave circuits locally with micrometer spatial resolution. We have compared and analyzed the LTLSM images of $J_{RF}(x,y)$ and $J_{IM}(x,y)$ that were simultaneously obtained in the meander strip Tl2212/LAO superconducting resonator by this technique. It has been shown that the sources of microwave nonlinearities may not be uniformly distributed across the film, and that IM PR is localized near the peaks in $J_{RF}(x,y)$, with only a very small fraction of the structure contributing to the global IM.

This work is supported in part by an NSF/GOALI grant DMR-0201261.